\documentclass[preprint,aps,notitlepage, nofootinbib,amsmath]{revtex4}
\usepackage{graphicx}% Include figure files
\usepackage{grffile}
\usepackage{color}
\usepackage{bm}% bold math
\usepackage{multirow}

\usepackage{}	% required for `\align' (yatex added)
\usepackage{}	% required for `\align' (yatex added)
\preprint{FQSP-26-1, YITP-26-45, OU-HET-1310}
\begin{document}

\title{Reply to ``Comment on “Chiral symmetry restoration, the eigenvalue density of the Dirac operator, and the axial U(1) anomaly at finite temperature”''}

\author{Sinya Aoki}
\affiliation{Fundamental Quantum Science Program (FQSP), TRIP Headquarter, RIKEN  Hirosawa 2-1, Wako, Saitama 351-0198, Japan
%Fundamental Quantum Science Program (FQSP), RIKEN, Hirosawa 2-1, Wako, Saitama 351-0198, Japan
}
\affiliation{Yukawa Institute for Theoretical Physics (YITP), Kyoto University, Kitashirakawa Oiwakecho, Sakyo-ku, Kyoto 606-8502, Japan}
\author{Hidenori Fukaya}
\affiliation{Department of physics, The University of Osaka, Toyonaka 560-0043, Japan}
\begin{abstract}
  We respond to the comment by Matteo Giordano \cite{Giordano}
on our article \cite{Aoki:2012yj}.
We find—and \cite{Giordano} itself acknowledges—that the proposed counterexamples
intended to refute our argument violate a crucial assumption of QCD at high temperatures,
namely that every gluonic observable is an analytic function of the squared quark mass, $m^2$.
%We further point out that the use of the density of states leads to a trivial circular argument.
%Even if the introduction of the density of states were justified,
%\cite{Giordano} relies on an incorrect assumption,
%which in turn leads to an incorrect conclusion regarding the $m^2$-analyticity of QCD.
We further point out a technical mistake found in \cite{Giordano}.
We conclude that the arguments presented in \cite{Giordano} are not valid.
\end{abstract}
\maketitle
\newpage
\section{$m^2$-analyticity of gluonic observable}

In \cite{Aoki:2012yj}, as in many other works on QCD,
we assumed that every gluonic observable in the chirally symmetric phase
is an analytic function of $m^2$ (which we refer to as $m^2$-analyticity for simplicity).
This assumption is based on the $m^2$-analyticity of the QCD partition function
above the critical temperature, together with the assumption that
this $m^2$-analyticity is preserved under the insertion of gluonic operators
or source terms into the action.

In Sec.~f of \cite{Giordano} the density-of-states function 
\begin{equation}
 \langle \mathcal{O}(A)^l \rangle = \int dx x^l \rho_{\mathcal{O}}(x,m),\;\;\; \rho_{\mathcal{O}}(x,m) = \langle\delta(x-\mathcal{O}(A))\rangle
\end{equation}
was introduced and gave several examples of $\rho_{\mathcal{O}}(x,m)$ for which
our argument in \cite{Aoki:2012yj} fails. 
But as the author of \cite{Giordano} himself admits, 
these ``counterexamples'' all break the above $m^2$-analyticity.

%\newpage
One of the examples raised in \cite{Giordano} is a simple Gaussian model:
\begin{equation}
\label{eq:Gaussian}
  \langle \delta(x-O(A)) \rangle = \frac{1}{\sqrt{2\pi}\sigma(m)}\exp\left[-\frac{(x-x_0(m))^2}{2\sigma(m)^2}\right]\theta(x).
\end{equation}
It is not difficult to confirm that for $\langle O(A)^\alpha \rangle$
with an non-integer $\alpha$ the above example 
gives a non-analytic power of $m^2$ in the $m\to 0$ limit.

%namely .
%The ``peculiar'' consequences that author of  \cite{Giordano} obtained is not because of
%the $m^2$-analyticity assumption but simply due to the wrong assumption (\ref{eq:wrongdelta})
%about the density of states.

In general, we can assume that the partition function for the massive system without 
spontaneously breaking of the symmetry, which is a highly nonlocal object,
is analytic in $m^2$ at $m=0$.
Any non-analyticity in $m$ reflects an infra-red divergence or criticality of the system.
As far as we know, there is no example of symmetric phase showing 
such infra-red critical behavior.
If the author of \cite{Giordano} claim that the $m^2$-analyticity on every observable 
is too strong, a clear counterexample should be given, where the system respects 
the full symmetry of the action, still showing the non-analyticity.

\section{Well-definedness of $P(m,A)$}

In  \cite{Aoki:2012yj} we argued that if 
\begin{equation}
 \lim_{m\to \infty}\frac{1}{m^k}\langle \mathcal{O}(A)^{l_0}\rangle =0,
\end{equation}
for some non-negative integers $l_0$ and $k$, then 
\begin{equation}
 \langle \mathcal{O}(A)^{l_0}\rangle = m^{k_0}\int DA P(m,A)\mathcal{O}(A)^{l_0},
\end{equation}
holds, where $k_0$ satisfies $k< k_0$ and $P(0,A)\neq0$.

We believe that the above two conditions are almost equivalent,
but it was argued in \cite{Giordano} that $P(0,A)$ is not necessarily well-defined.
Although the author of \cite{Giordano} himself admits that the equivalence is 
not crucial for the main discussion, we would like to 
show more details.

The definition of $\langle \mathcal{O}(A)^{l_0}\rangle$ at a finite quark mass $m$ 
is given by
\begin{align}
 \langle \mathcal{O}(A)^{l_0}\rangle &= \frac{1}{Z(m)}\int DA \det(D(A)+m)^{N_f}e^{-S_G(A)} \mathcal{O}(A)^{l_0},
\\
Z(m)&=\int DA \det(D(A)+m)^{N_f}e^{-S_G(A)},
\end{align}
where we assume that $N_f=2$, $D(A)$ is the massless Dirac operator, 
and $S_G(A)$ is the gluon part of the action.
With a characteristic function for the support of $\mathcal{O}(A)$ denoted by $\mathcal{S}(\mathcal{O})$,
\[
 \chi_{\mathcal{S}(\mathcal{O})}(A)=\left\{
\begin{array}{cc}
 1& \mbox{for}\; A \in \mathcal{S}(\mathcal{O}) \\
 0 & \mbox{otherwise}
\end{array}
\right.,
\]
the above expectation value can be written as
\begin{align}
 \langle \mathcal{O}(A)^{l_0}\rangle &= \frac{1}{Z(m)}\int DA \det(D(A)+m)^{N_f}e^{-S_G(A)} \chi_{\mathcal{S}(\mathcal{O})}(A) \mathcal{O}(A)^{l_0},
\end{align}
and we may identify 
\begin{equation}
 P(m,A) = \frac{1}{m^{k_0}Z(m)}\det(D(A)+m)^{N_f}e^{-S_G(A)} \chi_{\mathcal{S}(\mathcal{O})}(A).
\end{equation}
From the analyticity assumption with respect to $m$ above the critical temperature,
the functional integral of $P(m,A)\mathcal{O}(A)^{l_0}$ over $A$ should be an analytic function,
which satisfies all the required assumptions.

Note that $P(m,A)$ is given on a finite lattice with size $L$, for instance, with the overlap 
Dirac operator, which is well-defined at the level one can write down in a simulation code.
We do not see any problem in taking $L\to \infty$ limit of the above $P(m,A)$, while keeping $m>1/L$.

In \cite{Giordano}, an (unphysical) ill-defined example of $P(m,A)$ was suggested as
\[
P(m,A) = \frac{\Theta(m^2-|A|)}{m^{2l_0+2}}.
\]
But in this case, 
$\langle |A|^\alpha \rangle$ is non analytic with respect $m$ for non-integer $\alpha$,
which can be confirmed by the non-analytic contribution in $\frac{\partial}{\partial m^2}\langle |A|^\alpha \rangle$.
Therefore, this example is against the assumption of the $m^2$-analyticity of QCD in the symmetric phase.

\section{A wrong statement in \cite{Giordano}}

Sec.~h of \cite{Giordano} presents a discussion
that $m^2$-analyticity on nonlocal quantities is too strong,
taking an example with an operator
\begin{equation}
 \mathcal{O}_{\mathcal{L}}(A) = \frac{1}{V}\sum_x \mathcal{L}(A_x),
\end{equation}
where $x$ is averaged over whole lattice sites,
and consider
\begin{equation}
\label{eq:wrongdelta}
 \rho_{\mathcal{O}_{\mathcal{L}}}(x,m) = \delta(x-l_\mathcal{L}(m)),\;\;\; l_\mathcal{L}(m)=\langle \mathcal{O}_{\mathcal{L}}(A) \rangle.
\end{equation}
However, this equality is not valid, since $\langle \delta(x-O(A)) \rangle \neq \delta(x-\langle O(A)\rangle)$.
This can be easily checked by setting $\mathcal{L}(A_x)=\sum_{y}q(y)q(x)$ where $q(x)=\tilde{F}{F}$ is the 
topological charge density
and assume a simple Gaussian model for the $\theta$ dependence of the vacuum. 
If (\ref{eq:wrongdelta}) is assumed, we obtain a wrong result
$\langle \mathcal{O}_{\mathcal{L}}(A)^2 \rangle = \langle \mathcal{O}_{\mathcal{L}}(A)\rangle^2$,
while the true relation is $\langle \mathcal{O}_{\mathcal{L}}(A)^2 \rangle = 3\langle \mathcal{O}_{\mathcal{L}}(A)\rangle^2$.
It is well-known in general that the second cumulant (or the topological susceptibility) 
and fourth cumulant are independent.
The ``peculiar'' consequences obtained by the author of \cite{Giordano}
do not arise from the assumption of $m^2$-analyticity,
but rather from the incorrect assumption (\ref{eq:wrongdelta}) about the density of states.

\section{Summary}

We have found (and \cite{Giordano} itself admitted) that the counterexamples suggested in \cite{Giordano}
break a crucial assumption of QCD at high temperatures
that every gluonic observable is an analytic function of quark mass squared $m^2$.
%We have point out the problem in use of the density of states itself, 
%which  ended up with a trivial circular discussion.
We have argued that the well-definedness of $P(m,A)$ with the lattice regularization
using the overlap Dirac operator on a finite lattice.
We have also pointed out that the proposed form of $P(m,A)$ in \cite{Giordano}
breaks the $m^2$-analyticity.
Finally we found a wrong equality $\langle \delta(x-O(A)) \rangle = \delta(x-\langle O(A)\rangle)$ 
given in the argument of $m^2$-analyticity in \cite{Giordano}.
We conclude that the arguments by \cite{Giordano} are not valid.
%\newpage

%This work was supported in part by JSPS KAKENHI Grants No. JP22H00129, JP25K07283.
SA is supported by the RIKEN TRIP initiative. 
This work is supported in part by the JAPS Grant-in-Aid for Scientific Research (No. JP22H00129,  JP25K07283).


\begin{thebibliography}{99}
\bibitem{Giordano}
Matteo~Giordano, ``Comment on ``Chiral symmetry restoration, the eigenvalue density of
    the Dirac operator, and the axial U(1) anomaly at finite temperature'',''
preceding Comment, to appear in PRD.

%\cite{Aoki:2012yj}
\bibitem{Aoki:2012yj}
S.~Aoki, H.~Fukaya and Y.~Taniguchi,
``Chiral symmetry restoration, eigenvalue density of Dirac operator and axial U(1) anomaly at finite temperature,''
Phys. Rev. D \textbf{86}, 114512 (2012)
doi:10.1103/PhysRevD.86.114512
[arXiv:1209.2061 [hep-lat]].
%153 citations counted in INSPIRE as of 03 Dec 2025
\end{thebibliography}
\end{document}